\documentclass[12pt]{article}
\usepackage[english,german,french,polish]{babel}
\usepackage[T1]{fontenc}
\usepackage{amsfonts}

\begin{document}

\selectlanguage{english}

\baselineskip 0.73cm
\topmargin -0.4in
\oddsidemargin -0.1in

\let\ni=\noindent

\renewcommand{\thefootnote}{\fnsymbol{footnote}}

\newcommand{\SM}{Standard Model }

\newcommand{\SMo}{Standard-Model }

\pagestyle {plain}

\setcounter{page}{1}

%\pagestyle{empty}

%\addtocounter{equation}{+1}

~~~~~~
\pagestyle{empty}

\begin{flushright}
IFT-- 10/1
\end{flushright}

\vspace{0.4cm}

{\large\centerline{\bf Photonic portal to hidden sector}}

{\large\centerline{\bf and a parity-preserving option}}

\vspace{0.5cm}

{\centerline {\sc Wojciech Kr\'{o}likowski}}

\vspace{0.3cm}

{\centerline {\it Institute of Theoretical Physics, University of Warsaw}}

{\centerline {\it Ho\.{z}a 69, 00--681 Warszawa, ~Poland}}

\vspace{0.6cm}

{\centerline{\bf Abstract}}

\vspace{0.2cm}

In the case of previously proposed idea of photonic portal to hidden sector, the parity in this sector may be violated. We 
discuss here two new options within our model, where the parity is preserved. The first of them is not satisfactory, as not diplaying a full relativistic covariance. The second seems to be satisfactory.
 
\vspace{0.6cm}

\ni PACS numbers: 14.80.-j , 04.50.+h , 95.35.+d 

\vspace{0.6cm}

\ni January  2010

\vfill\eject

\pagestyle {plain}

\setcounter{page}{1}

\vspace{0.4cm}

\ni {\bf 1. Introduction}

\vspace{0.4cm} 
In previous papers [1,2], we have proposed a model of hidden sector of the Universe, consisting of sterile spin-1/2 Dirac fermions ("\,$\!$sterinos"), sterile spin-0 bosons ("\,$\!$sterons"), and sterile nongauge mediating bosons ("$A$ bosons") described by an antisymmetric-tensor field (of dimension one) weakly coupled to steron-photon pairs and, more obviously, to the antisterino-sterino pairs, 

%rownanie 1
\begin{equation}
- \frac{1}{2} \sqrt{\!f\,}\left(\varphi F_{\mu \nu} + \zeta \bar\psi \sigma_{\mu \nu} \psi \right) A^{\mu \nu}\,,
\end{equation}

\ni where $F_{\mu \nu} = \partial_\mu A_\nu - \partial_\nu A_\mu $ is the \SMo electromagnetic field (of dimension two), while $\sqrt{\!f\,}$ and $\sqrt{\!f\,}\,\zeta$ denote two dimensionless small coupling constants. Here, it is presumed that  $\varphi = <\!\!\varphi\!\!>_{\rm vac}\! + \,\varphi_{\rm ph}$ with a spontaneously nonzero vacuum expectation value
$<\!\!\varphi\!\!>_{\rm vac}\, \neq 0$. Such a coupling of photons to the hidden sector has been called "photonic portal" (to hidden sector). It provides a weak coupling between the hidden and \SMo sectors of the Universe. The photonic portal is an alternative to the popular "Higgs portal" \,(to hidden sector) [3].

The new interaction Lagrangian (1), together with the $A$-boson kinematic and \SMo electromagnetic Lagrangians, leads to the following field equations for  $F_{\mu \nu}$ and  $A_{\mu \nu}$ :

%rownanie 2
\begin{equation}
\partial^\nu \left[F_{\mu \nu} +  \sqrt{\!f\,}(<\!\!\varphi\!\!>_{\rm vac}\! + \,\varphi_{\rm ph}) A_{\mu \nu}\right] = -j_\mu \;\;,\;\; F_{\mu \nu} = \partial_\mu A_\nu - \partial_\nu A_\mu 
\end{equation}

\ni and

%rownanie 3
\begin{equation}
(\Box - M^2)A_{\mu \nu} = - \sqrt{\!f\,}\left[ (<\!\!\varphi\!\!>_{\rm vac}\! + \,\varphi_{\rm ph}) F_{\mu \nu} + \zeta \bar\psi \sigma_{\mu \nu} \psi\right]\,, 
\end{equation}

\ni  where $j_\mu$ denotes the \SMo electric current  and M stands for a mass scale of $A$ bosons, expected typically to be large.

The field equations (2), called~"\,$\!$supplemented Maxwell's equations", are modified due to the presence of hidden sector. This modification has a magnetic character, because the hidden-sector contribution to the total electric source-current  $j_\mu + \partial^\nu[ \sqrt{\!f\,}(<\!\!\varphi\!\!>_{\rm vac}\! + \,\varphi_{\rm ph}) A_{\mu \nu}]$ for the electromagnetic field $A_\mu$ is a four-divergence giving no contibution to the total electric charge $\int d^3x\{j_0 + \partial^k[ \sqrt{\!f\,}(<\!\!\varphi\!\!>_{\rm vac}\! + \,\varphi_{\rm ph}) A_{0 k}]\} = \int d^3x j_0 = Q$. In particular, it can be seen that the vacuum expectation value $<\!\!\varphi\!\!>_{\rm vac}\, \neq 0$ generates spontaneously a small sterino magnetic moment 

%rownanie 4%7
\begin{equation}
\mu_\psi = \frac{f \zeta}{2M^2}<\!\!\varphi\!\!>_{\rm vac} \,, 
\end{equation}

\ni though sterinos are electrically neutral. This is a consequence of an effective sterino magnetic interaction
 
%rownanie 5%6
\begin{equation}
- \mu_\psi \bar\psi \sigma_{\mu \nu} \psi F^{\mu \nu}  
\end{equation}

\ni appearing, when the low-momentum-transfer approximation

%rownanie 6%4
\begin{equation}
 A_{\mu \nu} \simeq \frac{\sqrt{\!f\,}\,\zeta}{M^2}\bar\psi \sigma_{\mu \nu} \psi 
\end{equation}

\ni effectively implied by Eq. (3) is used in the interaction (1) with $\varphi = <\!\!\varphi\!\!>_{\rm vac}\! + \,\varphi_{\rm ph}$. 

\vspace{0.4cm}  

{\bf 2. Option of independent field components for $A$ bosons}

\vspace{0.4cm}

In analogy with the familiar splitting of  $F_{\mu \nu}$ into $\vec{E}$ and $\vec{B}$, we can split the field $A_{\mu \nu}$ into the three-dimensional vector and axial fields $\vec{A}^{(E)}$ and $\vec{A}^{(B)}$ of spin 1 and parity  $-$ and +, respectively. Then,

%rownanie 7%10
\begin{equation} 
\left(A_{\mu \nu}\right) = \left(\begin{array}{rrrr} 0\;\;\;\;  & A^{(E)}_1 & A^{(E)}_2 & A^{(E)}_3 \\ -A^{(E)}_1 & 0\;\; \;\; & -A^{(B)}_3 & A^{(B)}_2 \\ -A^{(E)}_2 & A^{(B)}_3 & 0\;\; \;\; & -A^{(B)}_1 \\  -A^{(E)}_3 & -A^{(B)}_2 & A^{(B)}_1 & 0\;\; \;\; \end{array} \right) \,. 
\end{equation}

\ni Similarly, for the spin tensor $\sigma^{\mu \nu} = (i/2)[\gamma^\mu,\gamma^\nu]$ with $\vec{\alpha} = (\alpha_k) = (\gamma^0\gamma^k) = (i\sigma^{k 0})$ and $\vec{\sigma}=(\sigma_k) = \gamma_5\vec{\alpha} = (1/2)\left(\varepsilon_{k l m} \sigma^{l m}\right)\;(k=1,2,3)$, we get

%rownanie 8%12
\begin{equation} 
(\sigma^{\mu \nu}) = \left(\begin{array}{rrrr} 0\;\;  &i \alpha_1 & i \alpha_2 & i \alpha_3 \\ -i \alpha_1 & 0\;\; & \sigma_3 & -\sigma_2 \\ -i \alpha_2 & -\sigma_3 & 0\;\; & \sigma_1 \\  -i \alpha_3 & \sigma_2 & -\sigma_1 & 0\;\; \end{array} \right)\,. 
\end{equation}

\ni Then, the interaction (1) can be rewritten in the form

\vspace{0.1cm}

%rownanie 9%13
\begin{equation}
\left(\varphi \vec{E}- i\zeta \bar\psi \,\vec{\alpha} \,\psi \right)\cdot \vec{A}^{(E)} - \left(\varphi \vec{B} - \zeta \bar{\psi} \,\vec{\sigma}\psi \right)\!\cdot\!\vec{A}^{(B)} \,,
\end{equation}

\vspace{0.1cm}

\ni where $\varphi = <\!\!\varphi\!\!>_{\rm vac}\! + \,\varphi_{\rm ph}$ with $<\!\!\varphi\!\!>_{\rm vac}\, \neq 0$. Consequently, the first and second of supplemented Maxwell's equations (2) for photons can be split as follows:

\vspace{-0.2cm}

%rownanie 10%14
\begin{eqnarray} 
\vec{\partial} \times \left(\vec{B} + \sqrt{\!f\,}\varphi \vec{A}^{(B)}\right) = \,\;\partial_0 \!\left(\vec{E} + \sqrt{\!f\,}\varphi \vec{A}^{(E)}\right) + \vec{j} & , & \vec{\partial} \cdot \left(\vec{E} + \sqrt{\!f\,}\varphi \vec{A}^{(E)}\right) = j_0 \,, \nonumber \\ 
\vec{\partial} \times \vec{E}\;\;\; = -\partial_0 \vec{B}\;\;\; \;\;\;\;\; \;\;\;\;\; \;\;\;\;\; \;\;\;\;\; \;\;\;\;\; & , & \;\;\; \;\;\;\;\; \;\;\;\;\; \vec{\partial} \cdot \vec{B} = 0 
\end{eqnarray}

\ni and the field equation (3) for $A$ bosons as:

\vspace{-0.2cm}

%rownanie 11%14
\begin{eqnarray} 
(\Box - M^2)\vec{A}^{(E)} & = & - \sqrt{\!f\,}(\varphi \vec{E} - i \zeta \bar{\psi}\,\vec{\alpha}\,\psi )\,, \nonumber \\ 
(\Box - M^2)\vec{A}^{(B)} & = & - \sqrt{\!f\,}(\varphi \vec{B}\, - \,\zeta\bar{\psi}\, \vec{\sigma}\,\psi)\,,
\end{eqnarray}

\ni where $\varphi = <\!\!\varphi\!\!>_{vac}\! + \,\varphi_{\rm ph}$ with $<\!\!\varphi\!\!>_{\rm vac}\, \neq 0$. Here, $(j_\mu) = (j_0,-\vec{j}$) is the \SMo current ($\vec{E} = -\partial_0\vec{A} - \vec{\partial}A_0$ and $\vec{B} = \vec{\partial}\times \vec{A}$ with $(\partial_\mu) = (\partial_0,\vec{\partial})$ and $(A_\mu) = (A_0,-\vec{A})$). Note that the source-free Eqs. (10) are, of course, the ordinary source-free Maxwell's equations.

The sterile $A$ bosons described by the fields $\vec{A}^{\rm (E)}$ and $\vec{A}^{\rm (B)}$, when they propagate freely in space ($\sqrt{\!f\,} \rightarrow 0$), get the one-particle wave functions  

%rownanie 12%15
\begin{equation}
\vec{A}^{\,(E,B)}_{\vec{k}_A}(x) = \frac{1}{(2\pi)^{3/2}} \frac{1}{\sqrt{2 \omega_A}}\, \vec{e}^{\;(E,B)} e^{-i k_A\cdot x} \,,
\end{equation}

\ni where $k_A = (\omega_A, \vec{k}_A)$ with $\omega_A =\sqrt{\vec{k}_A^2 + M^2}$, while $\vec{e}^{\;(E,B)}$ are linear polarizations of $A^{(E)}$ and $A^{(B)}$ bosons [2].

If the fields $\vec{A}^{(E)}$ and $\vec{A}^{(B)}$ are independent (as can be in Eqs. (11)), then these polarizations form two triples of orthonormal versors,

%rownanie 13%16
\begin{equation}
\vec{e}_a^{\;(E,B)}\cdot \vec{e}_b^{\;(E,B)} = \delta_{a b}\;\; (a,b = 1,2,3) \;\;\;,\;\;\; \sum^3_{a=1} {e}_{a k}^{\,(E,B)} {e}_{a l}^{\,(E,B)} = \delta_{k l}\;\;(k,l = 1,2,3)
\end{equation}

\ni with $\vec{e}_a ^{\;(E,B)}\! = \!({e}_{a k}^{\;(E,B)}) \, (a=1,2,3,\; k=1,2,3)$ [2]. If the parity is preserved by the new weak interaction (1) or (9) in hidden sector, the polarizations $\vec{e}_a ^{\;(B)}\!$ ought to be axial vectors, while $\vec{e}_a ^{\;(E)}\!$ are polar vectors. In this case, the axial and polar vectors are $\vec{e}^{\;(B)}  = (-e_{23}, -e_{31}, -e_{12})$ and $\vec{e}^{\;(E)}  = (-e_{10}, -e_{20}, -e_{30})$, respectively, where $e_{\mu \nu} \,(\mu,\nu = 0,1,2,3)$ describe the antisymmetric polarization tensors appearing in the $A$-boson relativistic free wave function $A_{\mu \nu \vec{k}_A}(x)$  split into $\vec{A}^{\,(E,B)}_{\vec{k}_A}(x)$ given in Eqs. (12) (of course, there is a triplet of antisymmetric polarization tensors $e_{\mu \nu a} \;(a=1,2,3)$ split into two triplets $\vec{e}^{\,(E,B)}_a$ $(a=1,2,3))$. 

The axial $\vec{e}^{\,(B)}_a$, though defined carefully, are not practically realized for independent $\vec{A}^{\,(E)}$ and $\vec{A}^{\,(B)}$ fields. Therefore, in a real case, the field $\vec{A}^{\,(B)}$ may play a role of an effective polar vector of parity $-$ (like the field $\vec{A}^{\,(E)}$) and so, {\it the parity may be maximally violated} by the second term of coupling (9) in hidden sector [2]. This violation appears formally, when $\vec{e}^{\,(B)}_a$ are put polar (in spite of their original axial definition).

To be able to resign from such an option of indepedent field components for $A$ bosons, some Maxwell-type relations between the massive fields $\vec{A}^{(E)}$ and $\vec{A}^{(B)}$  (of dimension one) may be tentatively discussed as a new option, but it turns out to be not satisfactory (Section 3). Some relations of equivalence between the fields $\vec{A}^{(E)}$ and $\vec{A}^{(B)}$ may define still a different option, satisfactory this time (Section 4).

\vspace{0.4cm}

\ni {\bf 3. Option of Maxwell-type relations between field components for massive $A$~bosons} 

\vspace{0.4cm}

Consider two three-dimensional fields $\vec{X}^{(E)}$ and $\vec{X}^{(B)}$ of spin 1 and parity $-$ and +, respectively, satisfying the following set of first-order differential equations:

%rownanie 14
\begin{eqnarray} 
\vec{\partial} \times \vec{X}^{(B)} = \left(\partial_0 + iM\right) \vec{X}^{(E)}\! + \!\vec{\rho} \!& , & \vec{\partial} \cdot \vec{X}^{(E)} = \rho_{\,0} \;, \nonumber \\ 
\vec{\partial} \times \vec{X}^{(E)} = \left(-\partial_0 + i M\right)\vec{X}^{(B)}\!  \;\;\; & , &  \vec{\partial} \cdot \vec{X}^{(B)} = \;0 \;\;\,,
\end{eqnarray}

\ni where $(\rho_\mu) = (\rho_0, -\vec{\rho})$ is a four-vector fulfilling necessarily the condition

%rownanie 15
\begin{equation}
\vec{\partial} \cdot \vec{\rho} + (\partial_0 + i M)\rho_0 = 0
\end{equation}

\ni that would have the form of continuity equation if $M$ were zero (the {\bf div} operator is $\vec{\partial}\cdot $). Then, acting on the first and third Eq. (14) by the {\bf curl} operator $\vec{\partial} \times$ and applying the identity

%rownanie 16
\begin{equation}
\vec{\partial} \times(\vec{\partial} \times \vec{X}) = \vec{\partial}(\vec{\partial}\cdot \vec{X}) - \Delta \vec{X}
\end{equation}

\ni ($\Delta \equiv \vec{\partial}^{\,2}$), we conclude after combining both equations that

\vspace{-0.2cm}

%rownanie 17
\begin{eqnarray} 
(\Box - M^2)\vec{X}^{(E)} & = & (\partial_0 - i M)\vec{\rho} + \vec{\partial} \rho_0 \equiv -\vec{J}^{(E)} \,, \nonumber \\ 
(\Box - M^2)\vec{X}^{(B)} & = & - \vec{\partial} \times \vec{\rho} \equiv -\vec{J}^{(B)}  
\end{eqnarray}

\ni ($\Box \equiv \Delta - \partial_0^2$). We can see that any solution to Eqs. (14) satisfies also Eqs. (17) (but not necessarily {\it vice versa}), so the former are a sufficient condition for the latter.

Now, it is inferred from Eqs. (11) and (17) that, if the identities

\vspace{-0.2cm}

%rownanie 18
\begin{eqnarray} 
(-\partial_0 + i M)\vec{\rho} - \vec{\partial} \rho_0 & \equiv & \vec{J}^{(E)} \equiv \sqrt{\!f\,}\left(\varphi \vec{E} - i  \zeta\bar{\psi} \vec{\alpha} \psi \right) \;, \nonumber \\ 
\vec{\partial} \times \vec{\rho} & \equiv & \vec{J}^{(B)} \equiv \sqrt{\!f\,}\left(\varphi \vec{B} - \zeta\bar{\psi} \vec{\sigma} \psi \right) 
\end{eqnarray}

\ni were fulfilled, then our fields $\vec{A}^{(E)}$ and $\vec{A}^{(B)}$ might be used in place of $\vec{X}^{(E)}$ and $\vec{X}^{(B)}$ in Eqs. (14) and (17), where the former equations would be sufficient for the latter to hold (the latter would become Eqs. (11), being relativistic, as equivalent to the field equation (3) for $A_{\mu \nu}$). Then, $\vec{X}^{(E,B)} \equiv \vec{A}^{(E,B)}$ would have dimension one, while $\vec{\rho}$ and $\rho_0$ --- dimension two (and $\vec{J}^{(E,B)}$ --- dimension three). In this case, however, the lhs of identities (18) (together with Eq. (15)) would imply new relations  

\vspace{-0.2cm}

%rownanie 19
\begin{eqnarray} 
\vec{\partial} \times \vec{J}^{\,(B)} = (\partial_0 + i M) \vec{J}^{\,(E)} - (\Box - M^2)\vec{\rho} & , & \vec{\partial}\cdot \vec{J}^{\,(E)} = - (\Box - M^2)\rho_0 \;, \nonumber \\ 
\vec{\partial} \times  \vec{J}^{\,(E)} = (-\partial_0 + i M) \vec{J}^{\,(B)}\;\;\;\;\;\;\;\;\;\;\;\;\;\;\;\;\;\;\; & , &  \vec{\partial}\cdot \vec{J}^{\,(B)} = 0 
\end{eqnarray}

\ni which would be wrongly imposed by the rhs of Eqs. (18) on the independent fields $\vec{E},\,\vec{B}$ and $\varphi,\,\psi,\,\bar{\psi}$ (appearing then in $\vec{J}^{\,(E)}$ and $ \vec{J}^{\,(B)}$). This is so, since they should be related only through dynamical relationships provided by the field equations (following from the total Lagrangian).

In order to avoid these unwanted nondynamical relations, one may impose Eqs. (14) --- asymptotically ($\sqrt{\!f\,} \rightarrow 0$) and softly --- on the free wave functions (12) of massive $A$ bosons, writing

\vspace{-0.2cm}

%rownanie 20
\begin{eqnarray} 
\vec{\partial} \times \vec{A}^{(B)}_{\vec{k}_A} \;=\;\, (\partial_0 + i M) \vec{A}^{(E)}_{\vec{k}_A} & , & \vec{\partial}\cdot \vec{A}^{(E)}_{\vec{k}_A} = 0 \,, \nonumber \\ 
\vec{\partial} \times  \vec{A}^{(E)}_{\vec{k}_A} = \!(-\partial_0 + i M) \vec{A}^{(B)}_{\vec{k}_A} & , &  \vec{\partial}\cdot \vec{A}^{(B)}_{\vec{k}_A} = 0 
\end{eqnarray}

\vspace{-0.1cm}

\ni as a sufficient condition for the relativistic free one-particle wave equations{\footnote{An equivalent compact form of 
the asymptotic soft constraint (20) is

\vspace{-0.3cm}

$$ (\partial_\nu  + i M g_{0 \nu})A^{\mu \nu}_{ \vec{k}_A}  = 0 \;\;,\;\; (\partial_\nu - i M g_{0 \nu}) {\stackrel{\sim}{A}}\,\!^{\mu \nu}_{\vec{k}_A} = 0 ,$$ 

\vspace{-0.2cm}
 
\ni where ${\stackrel{\sim}{A}}\,\!^{\mu \nu} \!\equiv \!(1/2) \varepsilon^{\mu \nu \rho \sigma}A_{\rho \sigma}\;\, (\varepsilon^{0123} = 1)\,$ and so $A^{\mu \nu} \!\rightarrow\! {\stackrel{\sim}{A}}\,\!^{\mu \nu}$ when $\vec{A}^{(E)} \rightarrow \vec{A}^{(B)}$ and $\vec{A}^{(B)} \!\rightarrow\! -\vec{A}^{(E)}$ with $M \rightarrow -M$. Here, $g_{0 \nu}$ plays the role of a spurion for Lorentz boosts, spoiling relativistic covariance when $M \neq 0$. It dis\-appears in the relativistic wave equations (21) fulfilled necessarily, if the condition (20) is satisfied.}}

\vspace{-0.2cm}

%rownanie 21
\begin{equation}
(\Box - M^2) \vec{A}^{\,(E,B)}_{\vec{k}_A} = 0 \,. 
\end{equation}

With Eqs. (12), the asymptotic soft relations (20) show that

\vspace{-0.2cm}

%rownanie 22
\begin{equation}
\vec{k}_A \times \vec{e}^{\,(E)} = (\omega_A + M) \vec{e}^{\,(B)}\;,\; \vec{k}_A \cdot \vec{e}^{\,(E)} = 0 
\end{equation}

\ni and $\vec{k}_A \times \vec{e}^{\,(B)} \!=\! (-\omega_A \!+\! M) \vec{e}^{\,(E)},\, \vec{k}_A \!\cdot\! \vec{e}^{\,(B)} \!=\! 0$, what gives jointly $(-\vec{k}^2_A \!+\! \omega^2_A \!-\! M^2)\vec{e}^{\,(E,B)} = 0 $.{\footnote{In this case, the kinematics of $A$ bosons is relativistic, $k^2_A = M^2$, but the antisymmetric polarization tensors $e_{\mu \nu}$ are not relativistically covariant, since they satisfy the constraint 

\vspace{-0.3cm}

$$
(k_{A \nu} - M g_{0 \nu}) e^{\mu \nu} = 0 \;\;,\;\;(k_{A \nu} + M g_{0 \nu}) \stackrel{\sim}{e}\!^{\mu \nu} = 0
$$

\vspace{-0.1cm}

\ni involving the spurion $g_{0 \nu}$ when $M \neq 0$ (here, $\stackrel{\sim}{e}\!^{\mu \nu} = (1/2)\varepsilon^{\mu \nu \rho \sigma} e_{\rho \sigma}$, thus $e^{\mu \nu} \rightarrow \stackrel{\sim}{e}\!^{\mu \nu}$ when $\vec{e}^{\,(E)} \rightarrow \vec{e}^{\,(B)}$ and $\vec{e}^{\,(B)} \rightarrow -\vec{e}^{\,(E)}$ with $M \rightarrow -M)$. This compact form of constraint is equivalent to Eqs. (22) and two subsequent relations in the text. Note that the covariance of $e_{\mu \nu}$ appears in the limit of $M/\omega_A \rightarrow 0$.}} Hence, %311

\vspace{-0.2cm}

%rownanie 23
\begin{equation}
\vec{e}^{\,(B)} = \sqrt{\frac{\omega_A - M}{\omega_A + M}}\, \frac{\vec{k}_A }{|\vec{k}_A |}\times \vec{e}^{\,(E)} 
\end{equation}

\ni ($\omega_A = \sqrt{\vec{k}^2_A + M^2} $), both in right and lefthanded frame of reference. Thus, %w.311
$0< \vec{e}^{\,(B) 2}= (\omega_A - M)/(\omega_A + M) < 1$, if $\vec{e}^{\,(E) 2} = 1$ and $\omega_A > M > 0$. Note that $\vec{e}^{\,(B)} \rightarrow 0$, when $|\vec{k}_A |/M\rightarrow 0$ (as {\it e.g.} for an $A$ boson at rest). On the contrary, $\vec{e}^{\,(B)} \rightarrow (\vec{k}_A /|\vec{k}_A |) \times \vec{e}^{\,(E)} $, when $M/\omega_A \rightarrow 0($ {\it i.e.}, $M/|\vec{k}_A| \rightarrow 0$).

Concluding  our presentation of the Maxwell-type new option for massive $A$ bosons, we can say that now in their free wave functions there appear two not normalized to 1 axial vectors $\vec{e}^{\,(B)}$, dependent through the relations (23) on two (independent) polar vectors $\vec{e}^{\,(E)}$ forming together with $\vec{k}_A /|\vec{k}_A |$ a triple of orthonormal versors. So,

\vspace{-0.2cm}

%rownanie 24
\begin{eqnarray} 
\vec{e}^{\,(E)}_a\cdot \vec{e}^{\,(E)}_b = \delta_{a b} \;\;(a,b = 1,2) & , & \vec{e}^{\,(E)}_a \cdot \frac{\vec{k}_A}{|\vec{k}_A|} = 0 \;\;(a,b = 1,2) \,, \nonumber \\ 
\sum^2_{a =1}e^{\,(E)}_{a k} e^{\,(E)}_{a l} + \frac{k_{A k} k_{A l}}{\vec{k}^2_A} & = & \delta_{k l} \;\; (k,l = 1,2,3)
\end{eqnarray}

%\vspace{0.2cm}

\ni with $\vec{e}^{\,(E)}_a =  \left({e}^{\,(E)}_{a k}\right)\;(a = 1,2\;,\;k = 1,2,3)$. Then, two axial vectors $\vec{e}^{\,(B)}_{1,2}$ are parallel and antiparallel to the polar wectors $\vec{e}^{\,(E)}_{2,1}$, respectively, if $\vec{e}^{\,(E)}_1 \times \vec{e}^{\,(E)}_2 = \vec{k}_A /|\vec{k}_A |$  holds in the righthanded frame of reference. In such a new option, the coupling (9) {\it preserves the parity} in hidden sector, since $\vec{e}^{\,(B)}$ are practically realized as axial vectors, while $\vec{e}^{\,(E)}$ are polar vectors from the very beginning.

In this option, however, the asymptotic soft constraint (20) imposed on the free wave functions (12) of massive $A$ bosons is not relativistically covariant as far as the antisymmetric polarization tensors are concerned, although the kinematics is relativistic. Thus, the Maxwell-type new option turns out to be not satisfactory for massive $A$ bosons (when the full relativity does belong to our paradigm).

\vspace{0.4cm}

{\bf 4. Option of parallel $\vec{e}^{\,(E)}$ and $\vec{e}^{\,(B)}$ for $A$ bosons}

\vspace{0.4cm}

In contrast to the relations (23) between $\vec{e}_a^{\,(E)}$ and $\vec{e}_a^{\,(B)}\, (a= 1,2)$, other possible relations between them (now with $a=1,2,3$), namely   

%rownanie 25
\begin{equation} 
\vec{e}^{\,(B)}_{1,2,3} = \vec{e}^{\,(E)}_{2,3,1}\times \vec{e}^{\,(E)}_{3,1,2} = \left[ \left(\vec{e}_1^{\,(E)} \times \vec{e}_2^{\,(E)}\right) \cdot \vec{e}^{\,(E)}_3 \right]\vec{e}^{\,(E)}_{1,2,3} \;,
\end{equation} 

\ni are relativistically covariant. In fact, using Eqs. (25), we obtain for antisymmetric polarization tensors $e_{\mu \nu a}\;(a=1,2,3)$ the following forms

%rownanie 26
\begin{equation}  
e_{\mu \nu a} e^{\mu \nu}_a  \!=\! 2(\vec{e}_a^{\,(B) 2} \!-\! \vec{e}_a^{\,(E) 2}) \!=\! 2\left\{\!\left[\! \left(\vec{e}_1^{\,(E)} \times \vec{e}_2^{\,(E)}\right)\! \cdot\! \vec{e}^{\,(E)}_3 \right]^2 \!\!-\! 1\right\} \vec{e}^{\,(E)\,2}_a = 2\left[(\pm1)^2 - 1\right] = 0
\end{equation} 

\ni $(a = 1,2,3)$, being relativistically covariant in a trivial way, while in the case of relations (23) we get the forms

%rownanie 27
\begin{equation} 
e_{\mu \nu a} e^{\mu \nu}_a = 2(\vec{e}_a^{\,(B) 2} - \vec{e}_a^{\,(E) 2}) =  2 \left(\frac{\omega_A -M}{\omega_A +M} - 1\right) \vec{e}^{\,(E) 2}_a = - \frac{4M}{\omega_A + M}
\end{equation} 

\ni $(a = 1,2)$, violating the relativistic covariance when $M \neq 0$. When the relations (25) hold, Eqs. (13) are valid for $\vec{e}_a^{\,(E,B)}$ as previously for independent $\vec{e}_a^{\,(E)}$ and $\vec{e}_a^{\,(B)} \;(a=1,2,3)$.

Note that in the case of relations (25) the corresponding axial and polar vectors $\vec{e}_a^{\,(B)}$ and $\vec{e}_a^{\,(E)} \;(a=1,2,3)$, are parallel or antiparallel in the right or lefthanded frame of reference, respectively (in these frames, $\vec{e}_{2,3,1}^{\,(E)} \times \vec{e}_{3,1,2}^{\,(E)} = \pm\vec{e}_{1,2,3}^{\,(E)}$). An $A$ boson displays three independent polarizations that can be described by polar $\vec{e}_a^{\,(E)}$, since axial $\vec{e}_a^{\,(B)}$ are practically realized by means of relations (25) in terms of polar $\vec{e}_a^{\,(E)}$.

Thus, in  conclusion, the new option accepting the relations (25) between $\vec{e}_a^{\,(E)}$ and $\vec{e}_a^{\,(B)}\;(a=1,2,3)$ may be satisfactory in describing polarization of $A$ bosons. It is a scheme practically realizing axial $\vec{e}_a^{\,(B)}$ in terms of polar $\vec{e}_a^{\,(E)}$ and being relativistically covariant. In this option, {\it the parity is preserved} by the coupling (9) in hidden sector.

\vspace{2.5cm}

%\vfill\eject

{\centerline{\bf References}}

\vspace{0.4cm}

\baselineskip 0.73cm

{\everypar={\hangindent=0.65truecm}
\parindent=0pt\frenchspacing

{\everypar={\hangindent=0.65truecm}
\parindent=0pt\frenchspacing

[1]~W.~Kr\'{o}likowski, arXiv: 0803.2977 [{\tt hep--ph}]; {\it Acta Phys. Polon.} {\bf B 39}, 1881 (2008); {\it Acta Phys. Polon.} {\bf B 40}, 111 (2009); {\it Acta Phys. Polon.} {\bf B 40}, 2767 (2009). 

\vspace{0.2cm}

[2]~W.~Kr\'{o}likowski,  arXiv: 0909.2498 [{\tt hep--ph}];  arXiv: 0911.5614 [{\tt hep--ph}].

\vspace{0.2cm}

[3]~{\it Cf. e.g.} J. March-Russell, S.M. West, D. Cumberbath and D.~Hooper, {\it J. High Energy Phys.} {\bf 0807}, 058 (2008). 

\vspace{0.2cm}

\vfill\eject

\end{document}